\documentclass[twocolumn,english,aps,prb,twocolum,superscriptaddress,natbib,bibnotes,amsmath,amssymb,floatfix,groupedaddress,footinbib]{revtex4-2}
\usepackage[colorlinks=true,citecolor=blue,linkcolor=magenta]{hyperref}

\usepackage[addedmarkup=colored, authormarkupposition=left]{changes} 

\usepackage{soul}

\definechangesauthor[name={TJK}, color=red]{TJK}

\usepackage[utf8]{inputenc}
\usepackage[english]{babel}
\usepackage{amsmath,amsfonts,amssymb}
\usepackage[T1]{fontenc}
\usepackage{xurl}

\usepackage{amsmath}
\usepackage{siunitx}
\usepackage[version=4]{mhchem}
\newcommand{\dop}{{:}}

\usepackage{amsfonts}
\usepackage{amssymb}

\usepackage{epstopdf}
\usepackage{graphicx}
\graphicspath{{./Figures/}}

\begin{document}

\title{A fully hybrid integrated Erbium-based laser}

\author{	Yang Liu$^{1,2,\ast,\dag}$,
				Zheru Qiu$^{1,2,\ast}$, 
			    Xinru Ji$^{1,2}$, 
			    Andrea Bancora$^{1,2}$, 
			    Grigory Lihachev $^{1,2}$, 
			    Johann Riemensberger$^{1,2}$, 
			    Rui Ning Wang$^{1,2}$,
			    Andrey Voloshin$^{1,2}$,
				and Tobias J.  Kippenberg$^{1,2,\ddag}$}
\affiliation{
$^1$Institute of Physics, Swiss Federal Institute of Technology Lausanne (EPFL), CH-1015 Lausanne, Switzerland\\
$^2$Center for Quantum Science and Engineering,Swiss Federal Institute of Technology Lausanne (EPFL), CH-1015 Lausanne, Switzerland
}

\maketitle

\noindent\textbf{Erbium-doped fiber lasers \cite{barnes_high-quantum-efficiency_1989,suzuki_8_1989,chow_multiwavelength_1996} exhibit high coherence and low noise as required for applications in fiber optic sensing \cite{zadok_random-access_2012}, gyroscopes, LiDAR, and optical frequency metrology \cite{xu_recent_2013}. 
Endowing Erbium-based gain in photonic integrated circuits can provide a basis for miniaturizing low-noise fiber lasers to chip-scale form factor, and enable large-volume applications. 
Yet, while major progress has been made in the last decade on integrated lasers based on silicon photonics with III-V gain media \cite{tran2019ring,de2021iii,fan2020hybrid,morton2022integrated,guo2022hybrid,lihachev_low-noise_2022,xiang_high-performance_2021,li_reaching_2021, xiang_three-dimensional_2023}, the integration of Erbium lasers on chip has been compounded by large laser linewidth.
Recent advances in photonic integrated circuit-based high-power Erbium-doped amplifiers make a new class of rare-earth-ion-based lasers possible \cite{liu_photonic_2022}.
Here, we demonstrate a fully integrated chip-scale Erbium laser that achieves high power, narrow linewidth, frequency agility and the integration of a III-V pump laser. 
The laser circuit is based on an Erbium-implanted ultralow-loss silicon nitride (\ce{Si3N4}) photonic integrated circuit \cite{liu_photonic_2022}. 
This device achieves single-mode lasing with a free-running intrinsic linewidth of 50 Hz, a relative intensity noise of $<$-150 dBc/Hz at $>$10 MHz offset, and an output power up to 17~mW,  approaching the performance of fiber lasers \cite{fu_review_2017} and state-of-the-art semiconductor extended cavity lasers \cite{xiang_high-performance_2021,xiang_narrow-linewidth_2020,fan2020hybrid,li_reaching_2021}.
An intra-cavity microring-based Vernier filter enables wavelength tunability of $>$~40~nm within the C- and L-bands while attaining side mode suppression ratio (SMSR) of $>$ 70 dB, surpassing legacy fiber lasers in tuning and SMRS performance.
This new class of low-noise, tuneable Erbium waveguide laser could find applications in LiDAR \cite{kim_nanophotonics_2021}, microwave photonics \cite{liu_integrated_2020,marpaung_integrated_2019}, optical frequency synthesis \cite{spencer_optical-frequency_2018}, and free-space communications. 
Our approach is extendable to other wavelengths where rare-earth ions can provide gain, and more broadly, constitutes a novel way to photonic integrated circuit-based rare-earth-ion-doped lasers. 
}

\pretolerance=5000
Erbium-doped fiber lasers (EDFLs) \cite{barnes_high-quantum-efficiency_1989,suzuki_8_1989,chow_multiwavelength_1996} have become indispensable sources of high coherence laser light for distributed acoustic sensing \cite{soriano-amat_time-expanded_2021,zadok_random-access_2012}, optical gyroscopes, free-space optical transmission \cite{dix-matthews_point--point_2021}, optical frequency metrology\cite{xu_recent_2013}, and high-power laser machining \cite{nilsson_high_2010,jackson_towards_2012} and are considered the 'gold standard' of laser phase noise.
EDFLs exhibit many advantages such as all-fiberized cavities,  alignment-free components, 
and benefit from the advantageous Erbium-based gain properties including slow gain dynamics, temperature insensitivity, low amplification related noise figure,
lower spontaneous emission power coupled to oscillating modes than short semiconductor gain media \cite{siegman_lasers_1986,streifer_analysis_1982}, and excellent confinement of laser radiation for high beam quality.
These properties along with low phase noise have led to wide proliferation of Erbium-based fiber lasers in industrial applications.
Erbium ions can provide equally a basis for compact photonic integrated circuit-based lasers \cite{bradley_erbium-doped_2011} that can benefit from manufacturing at lower cost, smaller form factor and reduced susceptibility to environmental vibrations compared to fiber lasers.   
Prior efforts have been made to implement chip-based waveguide lasers using Erbium-doped materials such as \ce{Al2O3}\cite{ronn_ultra-high_2019,  mu_high-gain_2020}, \ce{TeO2}\cite{frankis_erbium-doped_2020}, \ce{LiNbO3}\cite{cai_erbium-doped_2022}, and Erbium silicate compounds \cite{wang_erbium_2018} as waveguide claddings or cores,  but the demonstrated laser intrinsic linewidth remained at the level of MHz \cite{li_monolithically_2018,belt_arrayed_2013, purnawirman_ultra-narrow-linewidth_2017,li_single-frequency_2021}, far above the sub-100-Hz linewidth achieved in commercial fiber lasers and state-of-the-art heterogeneously or hybrid integrated semiconductor-based lasers (Supplementary Note 1).
One major obstacle to realizing narrow-linewidth Erbium waveguide lasers is the challenge of integrating long and low-loss active waveguides ranging from tens of centimeters to meters---the lengths routinely deployed in fiber lasers to ensure low phase noise, single-frequency operation, and sufficient round-trip gain \cite{fu_review_2017}.

Here, we overcome this challenge and demonstrate hybrid integrated Erbium-doped waveguide lasers (EDWLs) using \ce{Si3N4} photonic integrated circuits that achieve narrow linewidth, frequency agility, high power, and the integration with pump lasers. 
Crucial to this advance are meter-scale-long Erbium-implanted silicon nitride (\ce{Er\dop Si3N4}) photonic integrated circuits that can provide $>$~30 dB net gain \cite{liu_photonic_2022} with $>$100 mW output power.
The \ce{Si3N4} photonic integrated circuit moreover exhibits absence of two-photon absorption in telecommunication bands\cite{liu_high-yield_2021}, radiation hardness for space compatibility, high power handling of up to tens of watts\cite{brasch_photonic_2016}, a lower temperature sensitivity than silicon, and low Brillouin scattering (a power-limiting factor in silica-based fiber lasers)\cite{gyger_observation_2020}.

\begin{figure*}[t!]
\centering
\includegraphics[width=\textwidth]{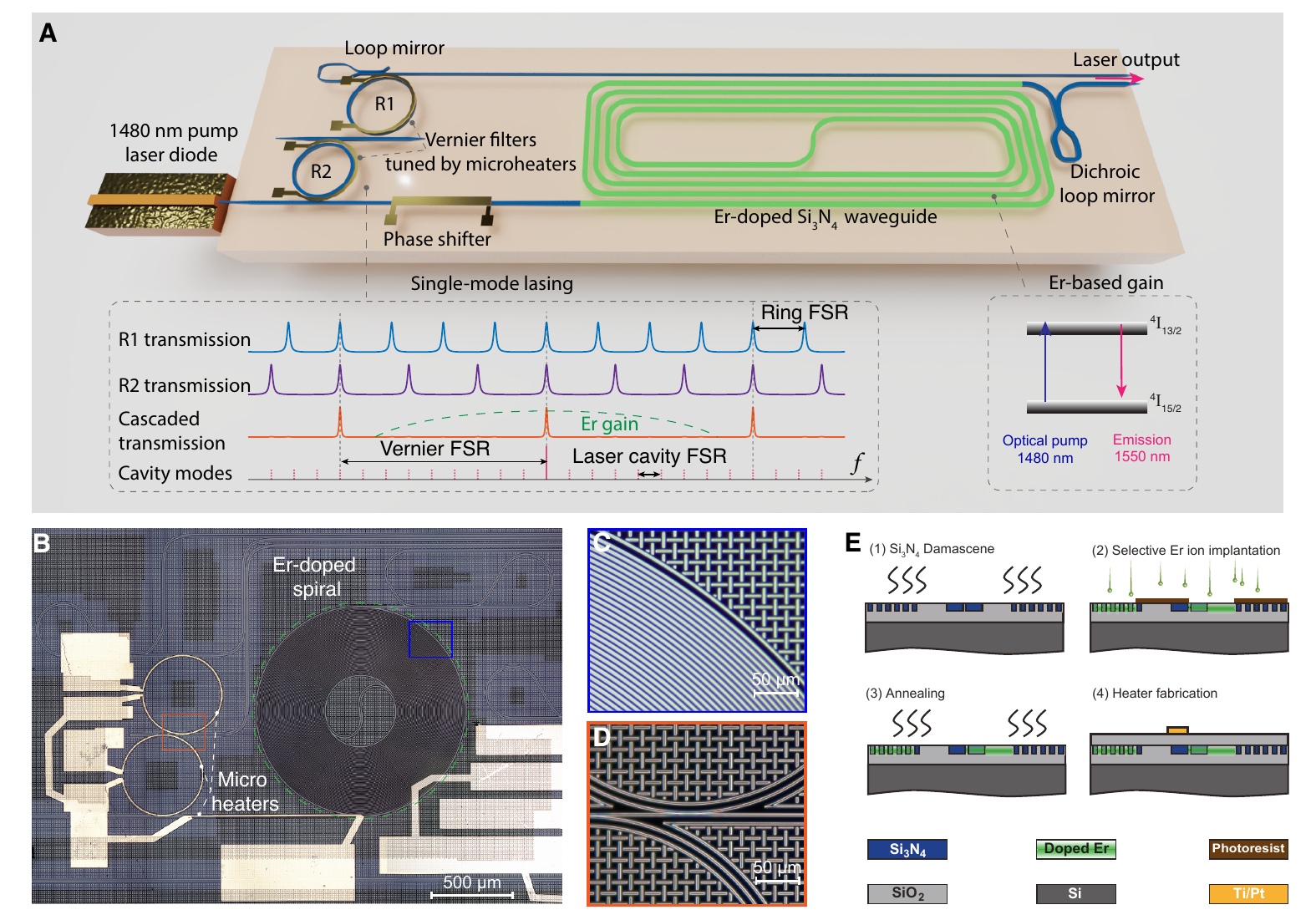}
\caption{
\footnotesize
\textbf{A hybrid integrated  \ce{Er\dop Si3N4} laser.} 
(\textbf{A}) Schematic of a hybrid integrated Vernier laser consisting of an Erbium-implanted silicon nitride \ce{Er\dop Si3N4} photonic integrated circuit and an edge-coupled III-V semiconductor pump laser diode.
The intra-cavity microring-based Vernier filter enables single-mode lasing operation within the Erbium-based gain bandwidth.
(\textbf{B}) Optical image of an \ce{Er\dop Si3N4} laser circuit integrated with micro heaters for wavelength and phase tuning. 
The green dashed circle indicates the Erbium-implanted gain spiral.
(\textbf{C}) Optical images of the Erbium-implanted spiral waveguides and (\textbf{D}) the coupling regime of the Vernier filter indicated by coloured boxes in (\textbf{B}).
(\textbf{E}) 
Fabrication process flow of the \ce{Er\dop Si3N4} photonic integrated circuit based on selective Erbium ion implantation. 
}
\label{Fig:1}
\end{figure*}

\section*{Results}
\subsection*{Hybrid integrated Erbium-based Vernier lasers}
The laser device is structured as a linear optical cavity with a spiral Erbium-doped gain waveguide and two reflectors formed by Sagnac loop mirrors at both ends (Fig.\ref{Fig:1}A). 
One dichroic loop mirror that consists of a dichroic directional coupler allows for laser reflection near 1550~nm and optical pump transmission near 1480~nm, and the other reflector deploys a short waveguide splitter for broadband reflection.
The optical pump can also be injected via a waveguide taper connected to a microring bus waveguide.
The laser device (Fig.\ref{Fig:1}B) exhibits a compact footprint of only 2 × 3 mm$^2$ with a densely-packed 0.2-m-long Erbium-doped \ce{Si3N4} spiral waveguide (Fig.\ref{Fig:1}C) with a cross section of \SI{0.7 \times 2.1}{\micro\metre\tothe{2}} .
A narrow-band intra-cavity Vernier filter designed to achieve sub-GHz 3~dB bandwidth and 5~THz FSR using two cascaded add-drop microring resonators (100~GHz FSRs with 2~GHz difference) (Fig.\ref{Fig:1}D) is deployed to ensure single-mode lasing operation with a small laser cavity mode spacing of ca. 200~MHz (Supplementary Note 2).
Integrated microheaters are used to align the Vernier filter peak transmission wavelength to a cavity longitudinal mode.
This integrated laser circuit was fabricated using the photonic Damascene process \cite{pfeiffer_photonic_2016}, followed by selective Erbium ion implantation, post annealing, and heater fabrication (Fig.\ref{Fig:1}E; see Methods and Supplementary Note 3). 

\begin{figure*}[t!]
\centering
\includegraphics[width=\textwidth]{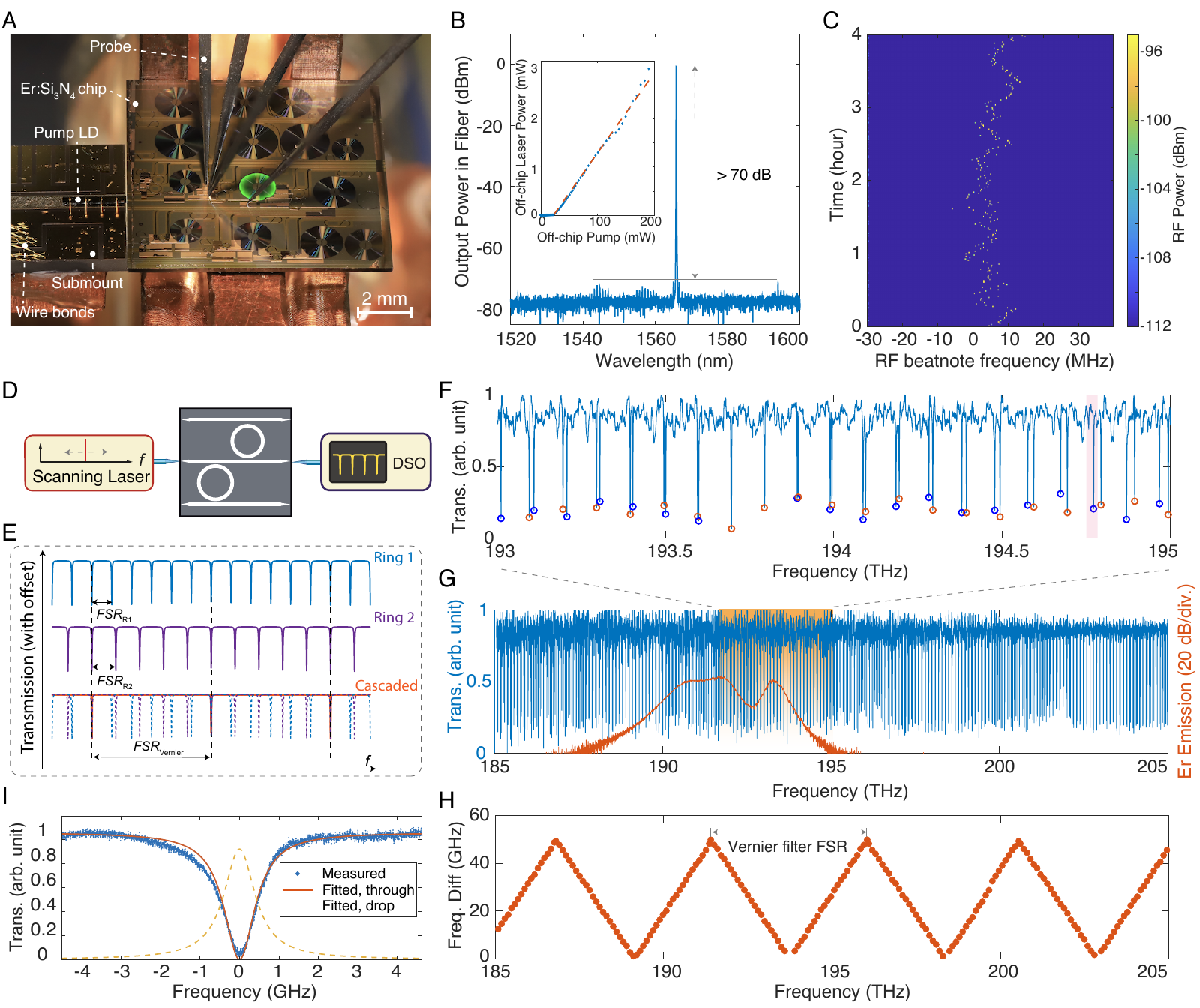}
\caption{ 
\footnotesize
\textbf{A hybrid integrated \ce{Er\dop Si3N4} vernier laser operated at single-mode lasing. } 
(\textbf{A}) Optical image of a hybrid integrated \ce{Er\dop Si3N4} Vernier laser edge-coupled with a pump laser diode chip (3SP Technologies, 1943 LCv1).  
Green luminescence was observed, stemming from the transition from higher-lying levels of excited Erbium ions to the ground state.
(\textbf{B}) Measured optical spectrum of single-mode lasing. 
The inset shows the output power as a function of the pump power.
(\textbf{C}) Measured time-frequency spectrogram of the heterodyne beatnote between the packaged EDWL and a fully-stabilized frequency comb (FC1500, Menlo Systems GmbH) over 4 hours.
(\textbf{D}) Experimental setup for Vernier filter characterization (device ID: \texttt{D85\_04\_F04\_C15\_V1}).
(\textbf{E}) Illustration of the Vernier effect by measuring the superposed resonances through the intermediate bus waveguide of the Vernier filter.
(\textbf{F}) Zoomed-in range of the measured transmission. 
Colored circles indicate the resonances of each microring.
(\textbf{G}) Wide-range transmission overlaid with the Erbium ion gain spectrum.
(\textbf{H}) Frequency spacing variation between adjacent ring resonances,  yielding a Vernier spacing of 4.65~THz, corresponding to 37.1~nm.
(\textbf{I}) The curve fitting of the measured through port transmission of the resonance indicated in (\textbf{F}), and the calculated filtering response at the drop port. 
}
\label{Fig:2}
\end{figure*}

To demonstrate a fully integrated EDWL, we performed photonic packaging via hybrid integration in a custom 14-pin butterfly package. 
The 1480 nm InP Fabry-P\'erot (FP) laser diode (LD) was edge coupled to one of the laser cavities on an \ce{Er\dop Si3N4} photonic integrated circuit (Fig.\ref{Fig:2}A), with simulated coupling loss of < 3 dB.
The laser output waveguide was end-coupled and glued with a cleaved UHNA-7 optical fiber spliced to a SMF-28 optical fiber pigtail, exhibiting 2.7~dB coupling loss at 1550~nm.
The pump LD,  a Peltier element, a thermistor, and all microheaters are connected to butterfly pins using wire bonding.
The integrated micro-heaters were used for the temperature control of the Vernier filter and the phase-shifter section to configure single-mode lasing and wavelength tuning.
The Erbium ions can be optically excited by the pump light emitted from the multi-longitudinal-mode pump LD ($>$4~nm spectral linewidth near 1480~nm), providing 1.9~ dB/cm of measured net gain coefficient \cite{liu_photonic_2022}.
The optical spectrum of the collected laser output shows a single-mode lasing operation with $>$~70 dB of side mode suppression ratio (SMSR) at 0.1 nm resolution bandwidth (Fig.\ref{Fig:2}B). 
This high 72-dB SMSR was made possible using the drop port of the the narrow passband intra-cavity Vernier filter, which can select the lasing mode and reject the broadband amplified spontaneous emission noise.
This record high SMSR surpasses what has been reported in integrated Erbium lasers, fiber lasers, and integrated semiconductor-based lasers (Supplementary Note 1), typically below 60~dB that is usually limited by intra-cavity filtering performance.
Conversely, this is challenging to implement in legacy fiber-based Erbium lasers where the filtering components based on long Bragg gratings can only offer several GHz wide passband with grating side lobes and lack of broadband wavelength tuning capability.
We observed an off-chip lasing threshold pump power of ca.  20~mW and an on-chip slope efficiency of 6.7~$\%$ when sweeping the pump power (Fig.\ref{Fig:2}B inset), which can be further optimized by reducing the coupling loss and the cavity loss.
The fully packaged laser showed a frequency drift of $<$ 20~MHz over 4 hours (Fig.\ref{Fig:2}C) when performing a heterodyne beatnote measurement with a fully-stabilized optical frequency comb indicating a good frequency stability due the monolithic nature of the laser comprised of both cavity and gain medium.
During a 24-hour test, this laser showed a frequency drift of $<$ 140~MHz without mode hops (Supplementary Note 4), representing a comparable long-term frequency stability as a commercial diode laser (Toptica CTL).

\subsection*{Single-mode lasing and wavelength tuning}
The use of photonic integrated circuits and Vernier structures (Fig.\ref{Fig:2}) enables to endow the integrated Erbium laser with broad wavelength tuning, a capability that bulk fiber lasers lack. 
We investigated the intra-cavity filtering properties by characterizing the optical transmission of the middle bus waveguide (Fig.\ref{Fig:2}D).
The measured transmission of the individual resonators used for the Vernier filter is shown in Fig.\ref{Fig:2}F and the designed 2 GHz FSR was experimentally attained (98~GHz and 100~GHz, respectively), leading to a measured Vernier filter FSR of 4.65~THz that corresponds to 37.1~nm span near 1550 nm wavelength (Fig.\ref{Fig:2}G,H).
Such a large Vernier FSR ensures the single-wavelength lasing within the Erbium emission wavelength range (Fig.\ref{Fig:2}G).
By overlapping the resonances from the two resonators, i.e. vanishing the frequency spacing (Fig.\ref{Fig:2}H), the lasing wavelength is determined.
By fitting the resonance linewidth near 194.8 THz (Fig.\ref{Fig:2}I), we obtain an external coupling rate $\kappa_{\mathrm{ex,0}}/2\pi=$~411~MHz (between the microring and the bus waveguide) and an intrinsic loss rate $\kappa_{\mathrm{0}}/2\pi=$~42.5~MHz. 
This strong over-coupled configuration ($\frac{\kappa_{\mathrm{ex}}}{\kappa_{\mathrm{0}}}>10$) can ensure that the Vernier filter \emph{simultaneously} achieves a narrow 3-dB passband bandwidth of 636~MHz and in principle a low insertion loss.
Such strong overcoupling can allow for low loss operation of the Vernier filter, which however in the current device was not attained.
The Vernier filter exhibits an insertion loss of -3.2~dB due to the parasitic loss induced by the coupling from the fundamental waveguide mode to higher order modes, which leads to a suboptimal coupling ideality \cite{pfeiffer_coupling_2017} of $I = 0.87$ (Supplementary Note 5).
 
\begin{figure*}[t!]
\centering
\includegraphics[width=\textwidth]{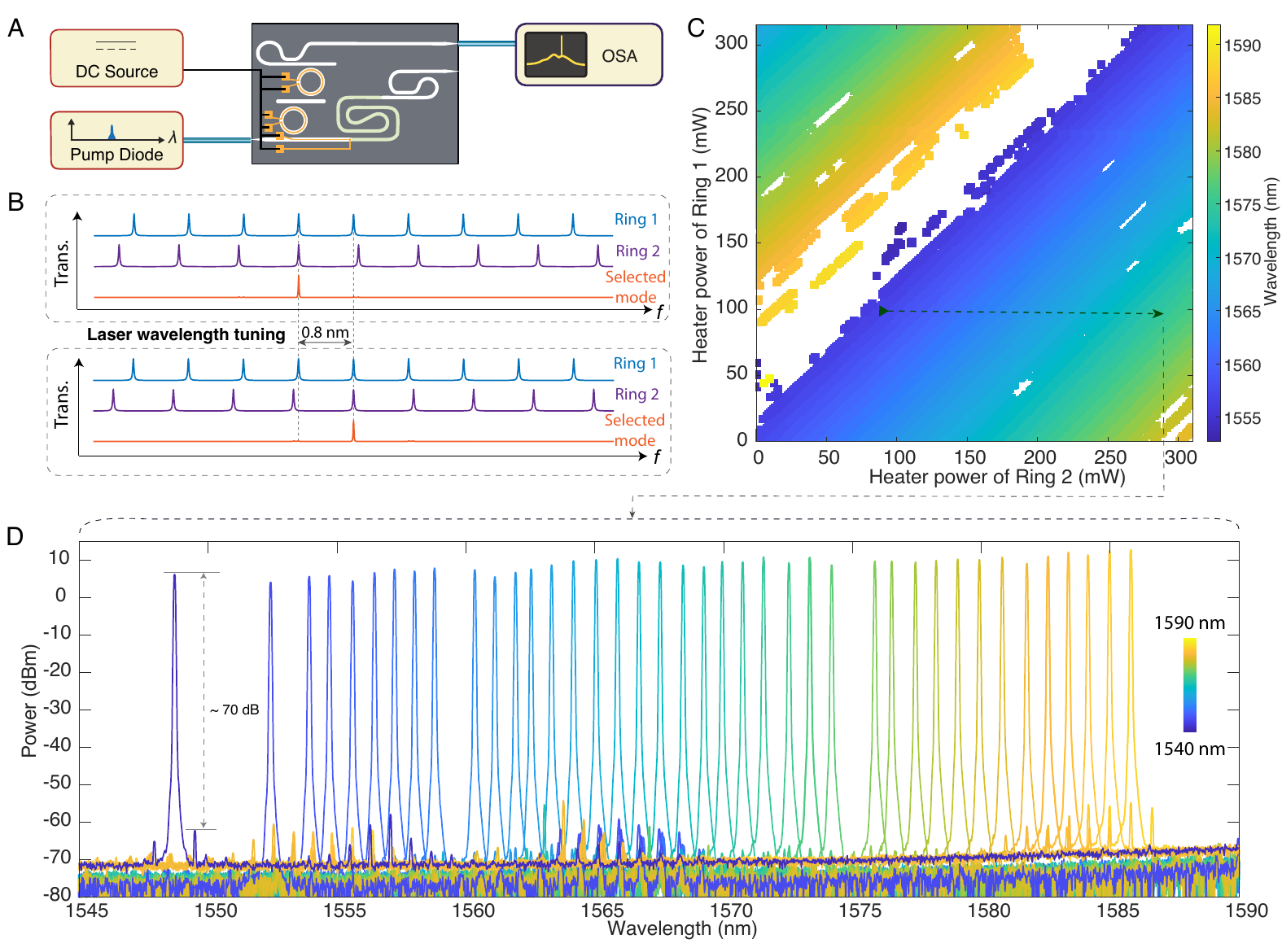}
\caption{
\footnotesize
\textbf{Demonstration of wideband tuning of the laser wavelength. }
(\textbf{A}) Experimental setup for laser wavelength tuning demonstration (device ID: \texttt{D85\_04\_F8\_C508\_VL2.0})
(\textbf{B}) Operating principle of the wavelength tuning of the vernier laser. The traces indicate the transmission of each over-coupled microresonator and the entire vernier filter, respectively. 
(\textbf{C}) Two dimensional laser wavelength tuning map, showing the wavelength of the predominant lasing mode as a function of the electrical power applied to the two micro heaters. 
The dashed schematically indicates the approach to coarse wavelength tuning. 
The white regions indicate that the expected laser emission was missing due to the microring resonance misalignment or a competing lasing mode when approaching the edge of the WDM filter transmission band.
(\textbf{D}) Measured optical spectra of single-mode lasing tuned over 40~nm wavelength range. 
The optical spectrum analyzer's resolution bandwidth is set to 0.1~nm. 
}
\label{Fig:3}
\end{figure*}

Next, we demonstrate the wavelength tunability (Fig. \ref{Fig:3}A). 
The coarse tuning of laser wavelength was carried out by switching the aligned resonance of two microresonators (Fig. \ref{Fig:3}B). 
The step size of ca. 0.8~nm was determined by the microring FSR.
Fine tuning of the wavelength can be achieved by simultaneously shifting the two resonators in the same direction and adjusting the phase shifter to align the corresponding cavity longitudinal mode (164~MHz spacing) to the Vernier filter passband.
Figure \ref{Fig:3}C shows the 2-dimensional (2D) laser wavelength tuning map when varying the electrical power applied to the microheaters.
From the recorded entire 2D map of wavelengths we selected the settings marked in Fig.\ref{Fig:3}C.
This allowed for continuous and deterministic tuning over the entire wavelength band from 1548.1~nm to 1585.8~nm, maintaining power of $>4~\mathrm{mW}$ and SMSR of $>70~\mathrm{dB}$ (Fig.\ref{Fig:3}D).
Such wavelength tunability cannot be achieved in conventional rare-earth-ion-doped fiber lasers without the use of free space etalon filters.
The wavelength tuning range was limited by the Vernier filter FSR and the wavelength-division multiplexing coupler transmission band (Supplementary Note 6).
During heater power scanning, we note that a few of wavelength tuning steps were missed due to the misalignment of microring resonances of the Vernier filter .
During tuning, the phase shifter was adjusted to maximize the output power at the desired mode.
A maximum fiber-coupled output power of ca. $17$~mW were measured at 1585~nm with 219~mW pump power.
Other competing lasing modes apart from the predominant lasing mode were observed when using high pump power, due to the fact that the large \ce{Si3N4} waveguide cross section allows for multiple transversal optical modes that can coincidently satisfy the lasing condition (Supplementary Note 8).

\begin{figure*}[t!]
\centering
\includegraphics[width=\textwidth]{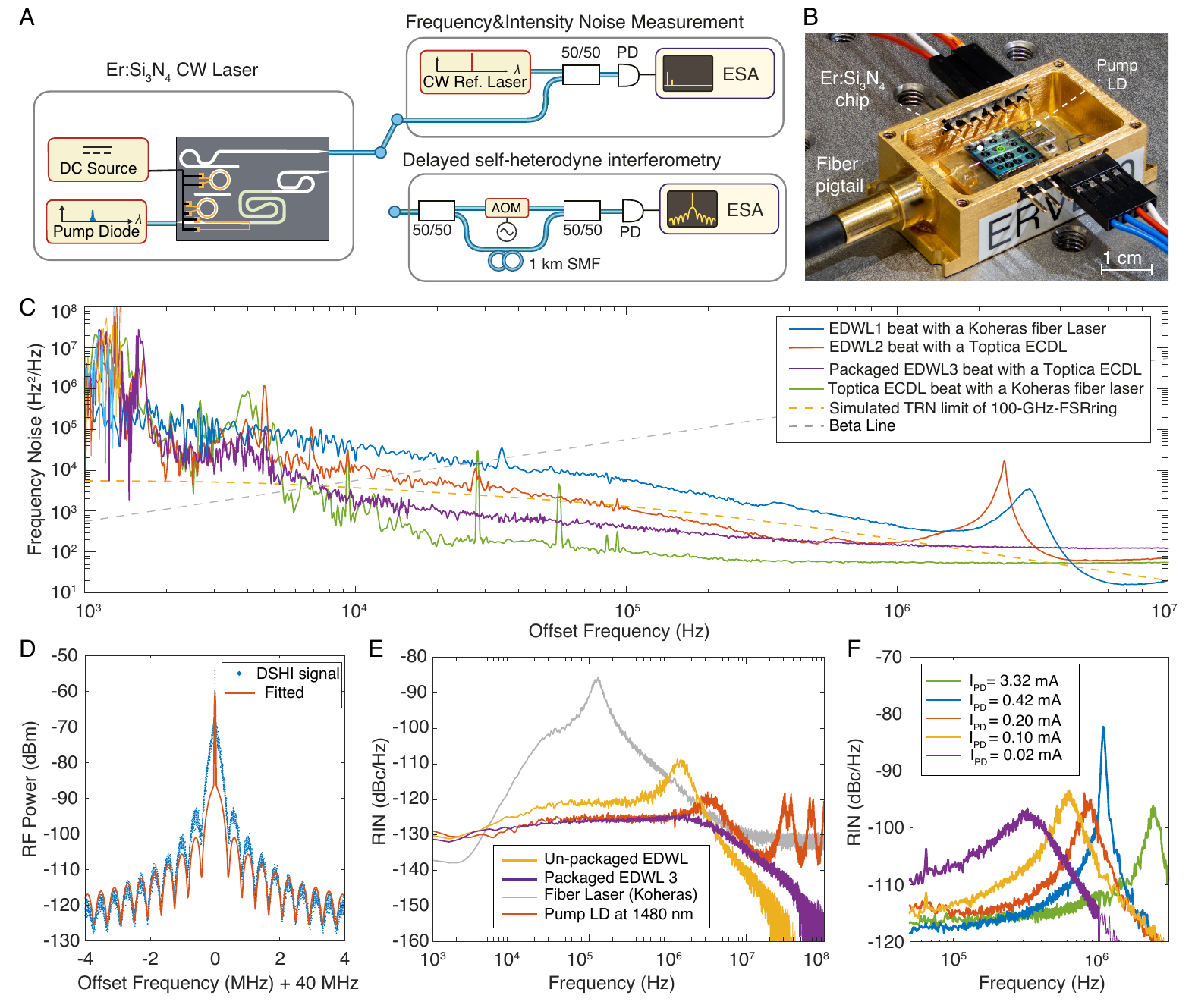}
\caption{
\footnotesize
\textbf{Laser noise properties and  the fully hybrid integration of an EDWL.}
(\textbf{A}) Experimental setups for the measurement of laser frequency noise, relative intensity noise, and intrinsic laser linewidth.
(\textbf{B}) Optical image of a fully hybrid integrated EDWL assembly.
(\textbf{C}) Measured laser frequency noise based on heterodyne detection with reference lasers. EDWL1 device ID: \texttt{D85\_04\_F8\_C508\_VL2.0}; EDWL2 device ID: \texttt{D85\_04\_F8\_C508\_VL2.1}; EDWL3 device ID: \texttt{D85\_04\_F8\_C508\_VL2.1}.
(\textbf{D}) Measured and fitted spectra of delayed self-heterodyne interferometric measurement for intrinsic laser linewidth investigation.
(\textbf{E}) Measured laser relative intensity noise (RIN) based on direct photodetetion.
(\textbf{F}) Relaxation oscillation peaks under varied pump power.
}

\label{Fig:4}
\end{figure*}
\subsection*{Frequency and intensity noise measurements}
To demonstrate the low noise features of the free-running EDWLs, we characterized the frequency noise,  the intrinsic laser linewidth, and the relative intensity noise (RIN), respectively (Fig. \ref{Fig:4}A).
Firstly,  a reference external cavity diode laser (free running Toptica CTL) was tuned close to the lasing wavelength near 1560~nm of an EDWL (not packaged) with ca.  3~mW output power for heterodyne photodetection.
The in-phase and quadrature components of the sampled beatnote time trace was processed using Welch's method \cite{welch_use_1967} to obtain the single-side power spectral density (PSD) of frequency noise $S_{\delta v}(f)$.
The frequency noise PSD (red line) reached a plateau of $h_0 = 62.0~\mathrm{Hz^{2}/Hz}$ at the offset frequency of 6 MHz, corresponding to a Lorentzian linewidth of $\pi h_0 = 194.8~\mathrm{Hz}$; this measured white noise floor was masked by the ECDL's white noise floor (Fig. \ref{Fig:4}C).
We also applied the delayed self-heterodyne interferometric measurement \cite{richter_linewidth_1986} to validate the intrinsic linewidth (Fig. \ref{Fig:4}D),  which generates a power spectrum of the autocorrelation of the laser line under sub-coherence condition (Supplementary Note 9).
In the offset frequency range from 10~kHz to 2.5~MHz where a relaxation oscillation peak was observed, the Erbium laser shows a higher frequency noise due to the laser cavity fluctuation caused by the pump laser noise transduction and the thermorefractive noise in the microresonator \cite{huang_thermorefractive_2019}.
The measured frequency noise at offset frequencies of $<$10~kHz was dominated by ECDL characteristic noise features \cite{liu_photonic_2020}.
We achieved a record low intrinsic linewidth (blue line) of $\pi h_0 = 50.1~\mathrm{Hz}$ ($h_0 = 15.9~\mathrm{Hz^{2}/Hz}$ ) in an Erbium waveguide laser with a higher output power of 10~mW, when beating against a low-noise Erbium fiber laser (Koheras Adjustik).
The fully packaged EDWL (Fig. \ref{Fig:4}B) with 2.8~mW output power shows a comparable intrinsic linewidth (purple line) and a lower frequency noise at the mid-range offset frequencies.
Using laser cavity designs with reduced cold cavity losses and increased mode area, hertz-linewidth EDWL can be feasibly achieved (Supplementary Note 10). 

The full width at half maximum (FWHM) of the integral linewidth associated with Gaussian contribution was obtained by integrating the frequency noise PSD from the inverse of measurement time (1/$T_0$) up to the frequency where $S_{\delta v}(f)$ intersects with the $\beta$-separation line $S_{\delta v}(f) = 8\ln(2)f/\pi^{2}$ (dashed line) \cite{di_domenico_simple_2010}.
With the integrated surface $A$, we obtained a minimum FWHM linewidth ($8\ln(2)A^{1/2}$) of the free-running EDWL is 82.2~kHz at 1~ms measurement time,
 which does not yet supersede a fiber laser, but is lower than 166.6~kHz of an ECDL (Toptica CTL) characterized as a reference laser for comparison.
For comparison, the commercial stabilized fiber-based laser shows 2.4~kHz of the FWHM linewidth at 1~ms measurement time.

Next, we show that the Erbium waveguide laser features a lower RIN compared to a commercial fiber laser (Koheras Adjustik, KOH45) (Supplementary Note 11).
The waveguide laser shows a RIN down to $-130~\mathrm{dBc/Hz}$ (yellow\& purple) at mid-range offset frequencies between 10~kHz and 1~MHz,  lower than the fiber laser RIN (grey) that has a PSD pole induced by relaxation oscillation (Fig. \ref{Fig:4}E).
The mid-range RIN was mainly limited by the pump laser RIN transduction which even contributed to an increased RIN by 5~ dB for the un-packaged EDWL.
The pump RIN noise transduction at frequency above $20~\mathrm{MHz}$ was suppressed due to the slow dynamics of Erbium ions.
The relaxation oscillation frequency can be calculated by $f_\mathrm{r} \approx \frac{1}{2\pi}\sqrt{\frac{P_\mathrm{cav}\kappa}{P_\mathrm{sat}\tau}}$ where  $P_\mathrm{sat}$ is saturation power of the gain medium,  $P_\mathrm{cav}$ is the laser cavity power,$\kappa$ is the cold cavity loss rate, and $\tau$ is the Erbium ion upper-state lifetime (Supplementary Note 12).
The waveguide laser RIN reduced to $< -155~\mathrm{dBc/Hz}$ at offset frequencies of $> 10$~MHz.
We observed that the relaxation oscillation frequency the waveguide laser varied from 0.3~MHz to 2.4~MHz when increasing the optical pump power (Fig. \ref{Fig:4}F),  which is higher than the case in the fiber laser (typically $<100~\mathrm{kHz}$ ). 
This higher relaxation oscillation frequency originates from the smaller saturation power and the shorter Erbium upper-state lifetime of  3.4~ms \cite{liu_photonic_2022}. 

\subsection*{Summary}
In summary, we have demonstrated a photonic integrated circuit-based Erbium laser that achievesd sub-100~Hz intrinsic linewidth, low RIN noise, $>$ 72~dB SMSR, and 40~nm wide wavelength tunability with power exceeding 10 mW.
The Erbium-doped waveguide lasers use foundry compatible silicon nitride waveguides, and have the potential to combine fiber-laser coherence with low size, weight, power and cost of integrated photonics. Such a laser may find application in existing applications such as coherent sensing,  and may equally provide a disruptive solution for emerging applications that require high volumes, such as lasers  for coherent FMCW LiDAR, or for coherent optical communications where iTLA (integrated tunable laser assembly) have been widely deployed, but fiber lasers' high coherence is increasingly demanded for advanced high-speed modulation formats while their use has been impeded by the high cost and large size. 
Co-doping other rare-earth ions such as ytterbium (emission at \SI{1.1}{\micro\metre}) and thulium (\SI{0.8}{\micro\metre},  \SI{1.45}{\micro\metre} and \SI{2.0}{\micro\metre}) will moreover allow access to other wavelengths. 
Looking to the future, the compatibility of silicon nitride with heterogeneously integrated thin-film lithium niobate\cite{churaev_heterogeneously_2021}, as well as piezoelectric thin films \cite{liu_monolithic_2020,lihachev_low-noise_2022}, and Erbium waveguide amplifiers \cite{liu_photonic_2022} provides the capability to create  fully-integrated high-speed, low-noise, high-power optical engines for LiDAR, long-haul optical coherent communications, and analog optical links.

\medskip
\begin{footnotesize}
\noindent \textbf{Funding Information}: This work was supported by the Air Force Office of Scientific Research (AFOSR) under Award No. FA9550-19-1-0250, and by contract W911NF2120248 (NINJA) from the Defense Advanced Research Projects Agency (DARPA),
Microsystems Technology Office (MTO). This work further supported by the EU H2020 research and innovation programme under grant No. 965124 (FEMTOCHIP), by the SNSF under grant no. 201923 (Ambizione), and by the Marie Sklodowska-Curie IF grant No. 898594 (CompADC) and grant No. 101033663 (RaMSoM).
\noindent \textbf{Acknowledgments}: 
Silicon nitride samples were fabricated in the EPFL Center of MicroNanoTechnology (CMi).  

\noindent \textbf{Author contributions}: 
Y.L.  and Z.Q.  performed the experiments.
Y.L.  carried out data analysis and simulations. 
Y.L.  and Z.Q.  designed \ce{Si3N4} waveguide laser chips.
X.J., G. L. and J.R. provided experimental supports. 
A.  B. and A. V. designed and performed the device packaging.
R.N.W., Z.Q.  and X.J.  fabricated the passive \ce{Si3N4} samples. 
Y.L.  wrote the manuscript with the assistance from Z.Q.  and the input from all co-authors. 
T.J.K supervised the project.

\noindent \textbf{Data Availability Statement}: 
The code and data used to produce the plots within this work will be released on the repository \texttt{Zenodo} upon publication of this preprint.

\noindent \textbf{Competing interests}
T.J.K. is a cofounder and shareholder of LiGenTec SA, a start-up company offering \ce{Si3N4} photonic integrated circuits as a foundry service. 


\end{footnotesize}

\renewcommand{\bibpreamble}{
$\ast$These authors contributed equally to this work.\\
$\dagger$\textcolor{magenta}{yang.lau@epfl.ch}\\
$\ddag$\textcolor{magenta}{tobias.kippenberg@epfl.ch}
}
\pretolerance=0
\bigskip
\bibliographystyle{apsrev4-2}
\bibliography{zotero_updated, library_additional}

\section*{Methods}
\subsection*{Device fabrication and ion implantation}
This ultralow loss \ce{Si3N4} photonic integrated laser circuit is fabricated using the photonic Damascene process \cite{pfeiffer_photonic_2016}. 
We applied selective Erbium ion implantation \cite{polman_optical_1991} (a total fluence of $1\times 10^{16}~\text{ions}~\text{cm}^{-2}$ at a maximum beam energy of 2~MeV) to the pre-fabricated passive \ce{Si3N4} photonic integrated circuits to endow the spiral waveguide with Erbium-based optical gain, while we kept other passive components un-doped by selectively masking a portion of the chip with photoresist (Fig.\ref{Fig:1}E and Supplementary Note 1). 
We achieved a high doping concentration of $3.25\times 10^{20}~\text{ions}~\text{cm}^{-3}$, more than one order of magnitude higher than that of conventional Erbium-doped fibers. 
This allows for high roundtrip net gain of 1.9~dB/cm (characterized from a 4.5-mm-long Erbium-doped waveguide) \cite{liu_photonic_2022}.
After ion implantation, we annealed the sample at \SI{1000}{\degreeCelsius} for one hour to activate the Erbium ions and heal implantation defects.
The optical gain is provided by the stimulated emission of erbium ions excited by an optical pump (Fig.\ref{Fig:1}A inset).
Micro-heaters were subsequently added atop the silica upper cladding after the ion implantation and post annealing processes.

\end{document}